\documentclass[letterpaper,twocolumn,english,aps,pra,showpacs,preprintnumbers]{revtex4}
\usepackage[T1]{fontenc}
\usepackage[latin9]{inputenc}
\usepackage{graphicx}
\usepackage{amssymb}
\usepackage{babel}
\usepackage{babel}
\usepackage{babel}

\makeatletter
\@ifundefined{textcolor}{}
{
 \definecolor{BLACK}{gray}{0}
 \definecolor{WHITE}{gray}{1}
 \definecolor{RED}{rgb}{1,0,0}
 \definecolor{GREEN}{rgb}{0,1,0}
 \definecolor{BLUE}{rgb}{0,0,1}
 \definecolor{CYAN}{cmyk}{1,0,0,0}
 \definecolor{MAGENTA}{cmyk}{0,1,0,0}
 \definecolor{YELLOW}{cmyk}{0,0,1,0}
 }
\makeatletter
\@ifundefined{textcolor}{}
{
 \definecolor{BLACK}{gray}{0}
 \definecolor{WHITE}{gray}{1}
 \definecolor{RED}{rgb}{1,0,0}
 \definecolor{GREEN}{rgb}{0,1,0}
 \definecolor{BLUE}{rgb}{0,0,1}
 \definecolor{CYAN}{cmyk}{1,0,0,0}
 \definecolor{MAGENTA}{cmyk}{0,1,0,0}
 \definecolor{YELLOW}{cmyk}{0,0,1,0}
 }
\makeatother
\makeatother

\begin{document}

\title{Sharp phase transitions in a small frustrated network of trapped ion spins}
\author{G.-D. Lin$^{1}$, C. Monroe$^{2}$, and L.-M. Duan$^{1}$}
\affiliation{$^{1}$Department of Physics and MCTP, University of Michigan, Ann Arbor,
Michigan 48109\\
$^{2}$Joint Quantum Institute, University of Maryland Department of Physics
and National Institute of Standards and Technology, College Park, MD 20742}
\date{\today }

\begin{abstract}
Sharp quantum phase transitions typically require a large system with many particles.
Here we show that for a frustrated fully-connected Ising spin network represented
by trapped atomic ions, the competition between different spin orders leads to rich phase transitions whose sharpness
scales exponentially with the number of spins.  This unusual finite-size scaling behavior opens
up the possibility of observing sharp quantum phase transitions in a system of just a few trapped ion spins.
\end{abstract}

\maketitle

Quantum simulators are motivated by the promise of gaining insight into many-body quantum systems
such as high-$T_C$ superconductors or complex arrangements of interacting spins.  Cold atomic systems
form a promising platform for quantum simulation, as the interactions between particles can be under great control.
A good example is a collection of trapped and laser-cooled atomic ions, each representing an
effective spin that can be made to interact with all the others by modulating the Coulomb interaction between ions.
By applying spin-dependent optical dipole forces it has been shown that a crystal of ions
provides an ideal platform to simulate intractable interacting spin models \cite{1,2}.
Following this proposal, recent experiments have simulated quantum magnetism with a few ions \cite{3,4,5,6}.
For three or more ions, the long-range coupling between the spins can provide frustrated interaction
patterns or competition between various spin orders and interesting quantum phase
diagrams \cite{5,6}. However, the observation of a quantum phase transition typically
requires a large system with many particles, as the width (sharpness) of
a quantum phase transition usually scales with $1/N$, the inverse of
the number of particles \cite{7}. With such slow finite-size scaling laws,
small networks of trapped ions realized in current experiments ($N\lesssim 20$)
are not expected to exhibit sharp transitions between distinct quantum phases.

In this paper, we show the surprising result that sharp phase transitions
can indeed be observed with just a few atomic ions. This is
due to unusual finite size scaling laws in this frustrated spin network,
where the sharpness of some phase transitions scales
exponentially instead of linearly with $1/N$. By controlling a single
experimental parameter that determines the pattern of spin-spin
couplings between the ions, we show that the expected ground state emerges from a delicate
compromise between the couplings. Frustration in the spin network leads to a variety
of spin orders, with the number of distinct phases increasing rapidly with the number of ions.
We construct the complete phase diagram for small spin networks realizable
with the current technology. The sharp phase transition is characterized in
detail with an explanation of its unusual finite-size scaling behavior.

We consider a small crystal of ions confined in a one-dimensional harmonic trap.
The spin states of the ions are represented by two internal states, referred as $\left\vert \uparrow
\right\rangle $ and $\left\vert \downarrow \right\rangle $, and the
effective spin-spin interaction between the ions is induced with off-resonant bichromatic
laser beams \cite{1,3,4,5}. The ion-laser coupling Hamiltonian, written in the
rotating frame, has the form $H=\sum_{n}\left[ \hbar \Omega \cos (\delta
kx_{n}+\mu t)\sigma _{n}^{z}+B\sigma _{n}^{x}\right] $ \cite{note}, where $%
\Omega$ is a Raman Rabi frequency, $\delta k$ is the wave vector
difference between the two Raman beams (which is assumed to be along the
radial direction $\hat{x}$), $\mu $ is the beatnote or detuning between the two laser beams,
$\sigma _{n}^{z}$ and $\sigma _{n}^{x}$ are Pauli matrices describing the spin of the $n$th ion,
and $B$ is an effective magnetic field induced by radiation that coherently flips the spins. In the
rotating frame, the radial coordinate $x_{n}$ is expanded in terms of the
transverse phonon modes $a_{k}$ as $x_{n}=\sum_{k}b_{n}^{k}\sqrt{\hbar
/(2m\omega _{k})}(a_{k}^{\dagger }e^{i\omega _{k}t}+a_{k}e^{-i\omega _{k}t})$%
, where $m$ is the atomic mass, $\omega _{k}$ is the eigenfrequency of the $%
k$th normal mode of the ion crystal, and $b_{n}^{k}$ is the eigenmode transformation matrix.
We use transverse phonon modes because they can more easily be
scaled up to large systems \cite{5,8}. Under the
Lamb-Dicke criterion $\eta_{n,k} \equiv b_{n}^{k}\delta k\sqrt{\hbar /(2m\omega _{k})}\ll 1$,
the Hamiltonian $H$ is simplified to
$H=-\hbar\Omega\sum_{nk}\eta_{n,k}\sin \left( \mu t\right)
\sigma _{n}^{z}(a_{k}^{\dagger }e^{i\omega _{k}t}+a_{k}e^{-i\omega _{k}t})%
 +B\sum_{n}\sigma _{n}^{x}$.

If we assume that the laser detuning $\mu$
is not resonant with any phonon mode with the condition $\left\vert \omega
_{k}-\mu \right\vert \gg \eta _{n,k}\Omega$ satisfied for all $n$ modes $k$,
the probability of exciting any phonon mode $\left\vert \Omega
\eta _{n,k}b_{n}^{k}/2\left( \omega _{k}-\mu \right) \right\vert ^{2}$ is
negligible. We can therefore adiabatically eliminate the phonon modes and
arrive at the following effective spin-spin coupling Hamiltonian \cite{6,9}%
\begin{equation}
H_{s}=\sum_{m,n}J_{mn}\sigma _{m}^{z}\sigma _{n}^{z}+B\sum_{n}\sigma
_{n}^{x},  \label{1}
\end{equation}%
where the coefficients%
\begin{equation}
J_{mn}=\frac{\left( \hbar \Omega \delta k\right) ^{2}}{2m}\sum_{k}\frac{%
b_{m}^{k}b_{n}^{k}}{\mu ^{2}-\omega _{k}^{2}}.
\end{equation}

This Ising Hamiltonian is a pillar of many-body physics, and its properties have been exhaustively studied
under various conditions \cite{7}.
For instance, the ground state of the Ising Hamiltonian is well understood when the coupling coefficients
$J_{mn}$ are uniform, or nonzero only for nearest neighbors. However, here we have an
extended Ising network where the coupling coefficients $J_{mn}$ are
inhomogeneous (both in magnitude and sign) and extend over long range
\cite{10}. The strong competition among these interaction terms (even with $B=0$) will generally lead
to highly frustrated ground states where individual bonds are compromised in
order to reach a global energy minimum.
For arbitrary coupling coefficients $J_{mn}$, the Hamiltonian (1) generally
belongs to the complexity class of NP-complete problems \cite{11}, meaning that
calculating attributes such as the system ground state becomes intractable when the system size is scaled up.

\begin{figure}[t]
\begin{centering}
\includegraphics[width=8cm]{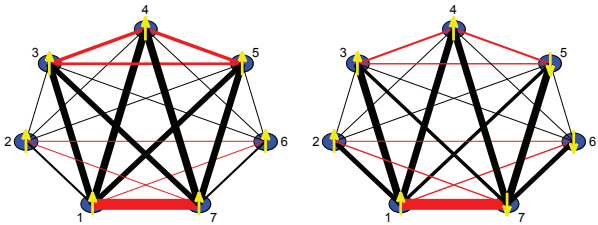}
\par\end{centering}
\caption{(Color online) Illustration of spin orders in two frustrated Ising networks
with competing long-range interaction for $N=7$ ions. \textit{Left panel}: Ferromagnetic (FM) order with
detuning $\tilde{\protect\mu}=5.1$; \textit{Right panel}: Kink order with $\tilde{\protect\mu}=5.3$.
The thickness of the edge in the graph represents the strength of each coupling. Positive (AFM)
spin couplings are indicated in red, while negative (FM) couplings are indicated in black.}
\label{fig:network}
\end{figure}

We consider the case where the coupling coefficients $J_{mn}$ are
controlled by a single experimental parameter, the laser detuning $\mu $ \cite{4,5,6}.
To determine $J_{mn}$ from any detuning $\mu $ with the formula (2), we first need
the normal mode transformation matrix $b_{n}^{k}$. This is obtained by finding the
equilibrium positions for a given number of ions in a harmonic trap and then
diagonalizing the Coulomb interaction Hamiltonian expanded about the ions' equilibrium
positions. With a single control parameter $\mu $, we are not able to program arbitrary
coupling coefficients $J_{mn}$. However, the interaction pattern is sufficiently
complex to allow frustrated ground state configurations and rich phase transitions. To illustrate
this we show in Fig. 1 a coupling pattern for $N=7$
ions and its associated ground state spin configuration for $B=0$. The
coupling pattern is represented by a graph where the color and the thickness
of each edge represents respectively the sign (ferromagnetic or
antiferromagnetic) and the magnitude of the coupling. In Fig. 1(a), we find
a ferromagnetically ordered ground state with all the spins pointing to
the same direction. However, in this ferromagnetic state, some of the bonds,
such as the strong antiferromagnetic bond between the ions $1$ and $7$, are
compromised, and due to this frustration, the ground-state spin
configuration is very sensitive to the strength of the coupling. If we adjust the
detuning $\mu $ by a small fraction of the trap frequency, the ferromagnetic bonds of the
ion pairs ($1,5$) and ($3,7$) are slightly weakened (see Fig. 1(b)) and the antiferromagnetic bond ($%
1,7$) dominates and flips the spin direction of the entire left (or right) half of
the ion crystal. This is a phase transition from ferromagnetic
order to a ``kink" order, with a kink in the spin direction between the 4th
and 5th ions counting from either the left or the right side.

\begin{figure*}[t]
\begin{centering}
\includegraphics[width=13cm]{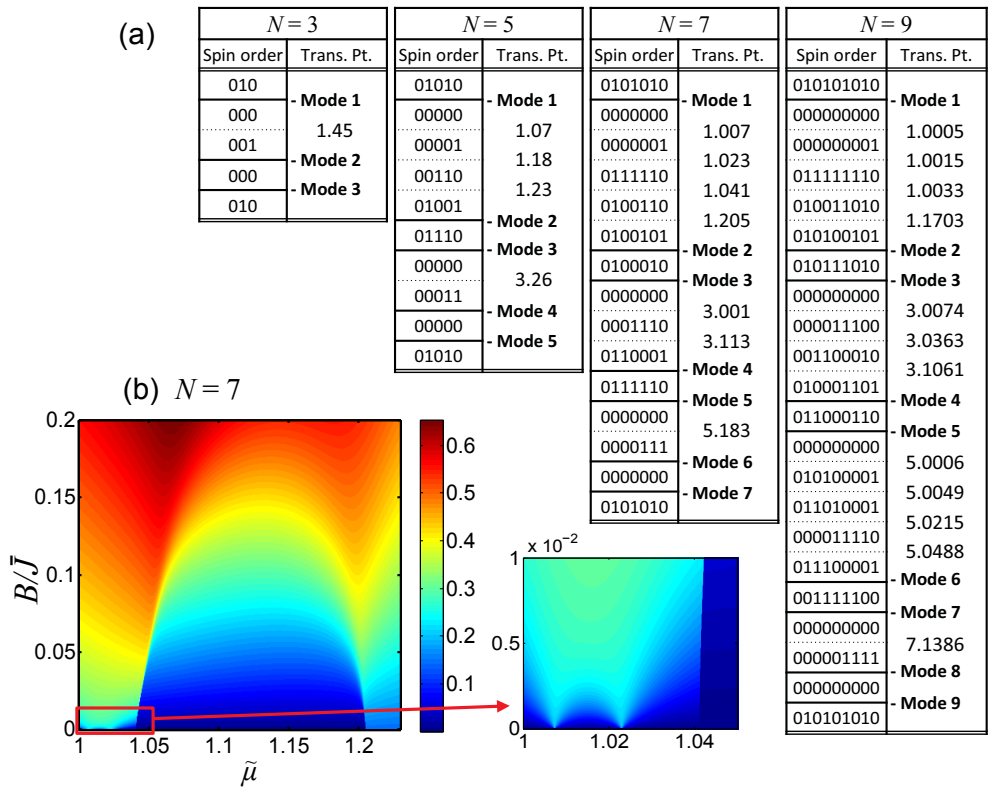}
\par\end{centering}
\caption{(Color online) (a) Ground-state phase diagram at $B=0$ characterized by
the corresponding spin orders for $N=3, 5, 7 ,9$ ions. The transition point is denoted by the
re-scaled detuning $\tilde{\mu}$. (b) Average polarization $\bigl\langle\sum_{n}\sigma _{n}^{x}\bigr\rangle/N$ for $N=7$ ions
at finite fields $B$. The right figure is a closeup near the critical point. }
\label{fig:nodd_table}
\end{figure*}

To show the rich phase diagram for this system, in Fig. 2(a) we list all
different spin phases at $B=0$ for a small Ising network with $3$, $5$, $7$
and $9$ ions. For an odd number of ions, the phase diagram is more interesting
and features a larger variety of spin orders, because the left-right reflection
symmetry in a linear ion crystal can be spontaneously broken.
In Fig. 2, for convenience, we denote the laser detuning $\mu $
with a re-scaled dimensionless parameter $\tilde{\mu}$ by labeling the
phonon-mode eigen-frequencies (from lowest to highest) with integers (from $1
$ to $N$ for an $N$-ion crystal) and using a linear scaling between two
integers to denote detuning located between the corresponding two modes. For
instance, in this notation $\tilde{\mu}=2.75$ means the physical detuning $%
\mu =\omega _{2}+0.75(\omega _{3}-\omega _{2})$. Each phase is characterized
by a spin order (denoted with a binary string where $0$ and $1$ correspond
to $\uparrow $ and $\downarrow $ spin respectively) which gives one of the
ground state spin configurations. The Ising Hamiltonian (1) features a
reflection symmetry and an intrinsic $Z_{2}$ symmetry with respect to a
global spin flip. The spin order breaks the Ising symmetry, so
each phase is at least two-fold degenerate. If the spin order also breaks
the reflection symmetry, the corresponding ground state is $4$-fold
degenerate. For instance, for the phase denoted by the spin order $01001$,
the four degenerate ground states are $\bigl|\downarrow \uparrow \downarrow
\downarrow \uparrow \bigr\rangle$, $\bigl|\uparrow \downarrow \uparrow
\uparrow \downarrow \bigr\rangle$, $\bigl|\uparrow \downarrow \downarrow
\uparrow \downarrow \bigr\rangle$, and $\bigl|\downarrow \uparrow \uparrow
\downarrow \uparrow \bigr\rangle$. When the re-scaled detuning $\tilde{\mu}$
crosses an integer (a phonon mode), the spin order changes as expected,
but this is not a conventional phase transition as the parameters $J_{mn}$
change discontinuously in the Hamiltonian (1). However, when $\tilde{\mu}$
varies within two adjacent integers, all the parameters $J_{mn}$ are
analytic functions of $\tilde{\mu}$, yet the spin order can still change
abruptly, signaling a phase transition. The frequency of this type of
inter-mode phase transition increases rapidly with the ion number: there is
one such transition for a three-ion chain and $12$ such transitions in a
nine-ion crystal. Another notable feature from Fig. 2(a) is that there is
typically no phase transition when $\tilde{\mu}$ varies from an even
mode ($2k$) to an odd mode ($2k+1$, $k=0,1$,$\cdots ,\left( N-1\right) /2$).
In such regions, the spin order has a reflection symmetry. This suggests
that a spin order with reflection symmetry may be more stable in
energy and does not easily yield to other spin configurations. This
observation is consistent with the fact that for an even number of ions,
there are much fewer inter-mode phase transitions, as the spin order in these cases
has a reflection symmetry.

As we add a transverse $B$\ field to the Hamiltonian, the spins will gradually
become polarized along the $x$-direction along $B$. In Fig. 2(b), we plot the average
polarization $\bigl\langle\sum_{n}\sigma _{n}^{x}\bigr\rangle/N$ as a
function of the field $B$ (in the unit of the average $J_{mn}$ defined by $%
\bar{J}\equiv \sqrt{\sum_{m\neq n}|J_{mn}|^{2}/[N(N-1)]}$) for $N=7$ ions in
a small region of the detuning $\tilde{\mu}$. We find that the
system is easily polarized if it lies at the critical point between two
different spin orders given by the Ising couplings. But near the center of a spin phase, the spin order is more
robust and can persist under a finite $B$, eventually yielding to the
polarized phase as $B$ increases through the Ising-type transition (which
becomes a broad crossover for this finite system).

\begin{figure*}[t]
\begin{centering}
\includegraphics[width=15.5cm]{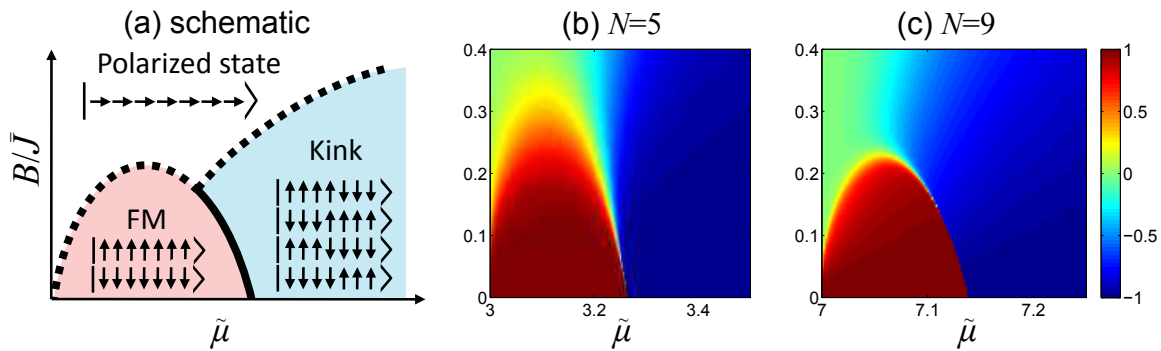}
\par\end{centering}
\caption{(Color online) (a) The schematic phase diagrams for odd numbers of
ions in the region with detuning $N-2<\tilde{\mu}<N-1$. For a small number of ions,
the solid line represents a sharp transition, whereas the dashed lines represent a continuous crossover to the polarized state.
(b,c) The calculated theoretical phase diagrams for (b) $%
N=5$ and (c) $N=9$ ions. Color shows the order parameter defined by $P_{FM}-P_{K}$,
where $P_{FM}$ and $P_{K}$ are the projection probabilities of the
ground state of the system to the Hilbert subspace with the ferromagnetic and the kink orders, respectively
(the basis-vectors of the corresponding subspaces are indicated in Fig. (a)).}
\label{fig:magic_pt}
\end{figure*}

With $B=0$, the transition between different spin orders is sharp as it is
characterized by a level crossing for the ground state of the Hamiltonian
(1). When we turn on a finite $B$ field, the system shows only avoided level
crossings in its ground state, and we expect that the sharp phase transition
at $B=0$ to be replaced by a broad crossover for this small system,
similar to the Ising-type of transition discussed above. Interestingly, this
is not always the case. Even at a finite $B$, some transitions
remain very sharp in this small system, and the sharpness increases
exponentially with the number of particles. This provides an unusual
finite-size scaling behavior, distinctively different than the polynomial
finite-size scaling of the transition width that one typically sees in
conventional phase transitions \cite{7}. (For instance, the conventional
Ising transition displays a linear scaling law \cite{7}.) To characterize
this unusual finite-size scaling behavior, we look at a particular example:
for $N$ ions, there is a unique spin phase transition in the region with
detuning $N-2<\tilde{\mu}<N-1$ when $N$ is odd. A schematic phase diagram
for this region is shown in Fig. 3(a). At $B=0$, we have a ferromagnetic
phase on the left side which is doubly degenerated and a kink phase on the
right side which is $4$-fold degenerate. At finite $B$, these two spin
orders remain robust in a range of $B$, before eventually yielding
to a polarized phase for a large $B$ field through a crossover.
The transition between the ferromagnetic and the kink phases remain sharp
for a finite $B$, and this sharpness increases rapidly with the ion number,
as one can see from Fig. 3(b) and 3(c). With a moderate increase of the ion
number from $5$ to $9$, the transition between the ferromagnetic and the
kink phase becomes almost infinitely sharp. It is also interesting to note
that the transition line between these two phases has a slope with the $B$
axis, so one can cross this phase transition by tuning either the detuning $%
\tilde{\mu}$ or the field $B$.

\begin{figure}[tbp]
\begin{centering}
\includegraphics[width=8cm]{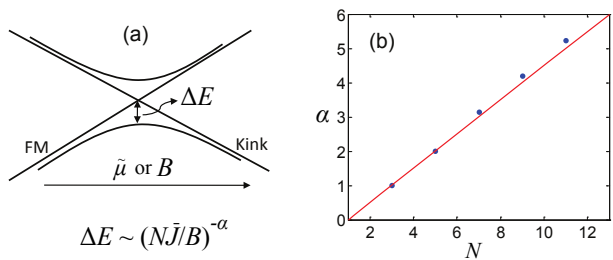}
\par\end{centering}
\caption{(a) The structure of the lowest four energy levels around the
sharp ferromagnetic-kink phase transition. The two lowest states have a ferromagnetic (kink)
order on the left (right) side. The transition width $W$ is proportional to the energy gap $\Delta E$, and
$\Delta E$ is fitted by an exponential. (b) The exponent $\protect\alpha $ is a function of the
ion number $N$, where the solid line is a linear fit $\alpha \simeq \left(
N-1\right) /2$. }
\label{fig:gap}
\end{figure}

We have characterized the transition width between the ferromagnetic phase
and the kink phase, and find this transition width indeed shrinks
exponentially with the number of ions. To see this, we show
the four lowest levels across this phase transition when we tune either $%
\tilde{\mu}$\ or $B$ (see Fig. 4a). While the lowest state is
a smooth function of $\tilde{\mu}$ at a finite $B$, the second and third
states feature a level crossing. From Fig. 4a, the transition
width $W$ is proportional to the energy gap $\Delta E$. We calculate the
energy gap $\Delta E$ as a function of the ion number $N$, and find it is
well fit by $\Delta E\propto \Bigl(B/N\bar{J}\Bigr)^{\alpha }$ for $B\ll N%
\bar{J}$, where the exponent $\alpha $ has a roughly linear dependence
with $N$ as shown in Fig. 4b and can be fit by $\alpha \simeq \left(
N-1\right) /2$. This formula shows that the energy gap $\Delta E$, and thus
also the transition width $W$, shrinks exponentially with the ion number,
which explains why the transition is so sharp for a small system. The
exponential shrinking of the energy gap $\Delta E$ can be intuitively
understood as follows: when $B\ll N\bar{J}$ we can treat the term $%
B\sum_{n}\sigma _{n}^{x}$ as a perturbation in the Hamiltonian (1). For each
application of $B\sum_{n}\sigma _{n}^{x}$, we can only flip the direction of
one spin. As the ferromagnetic state and the kink state have $\left(
N-1\right) /2$ spins taking opposite directions, the two states need to be
connected through $\left( N-1\right) /2$th order perturbation, and thus the
energy gap is proportional to $\left( B/N\bar{J}\right) ^{\left( N-1\right)
/2}$.

In summary, we have shown that laser induced magnetic coupling between
trapped ions realizes a frustrated Ising spin network with competing long
range interactions, giving rise to rich phase diagrams for the
ground state. Some of the phase transitions in this system are characterized
by a unusual finite size scaling, where the transition width scales down
exponentially with the number of ions. This exponential finite size scaling
lads to sharp phase transitions for a small system even with just a few
ions, as one can realize now in the lab.

This work was supported by the DARPA OLE program, the IARPA MQCO program, 
the AFOSR and the ARO MURI programs, the NSF PIF Program, and the NSF Center at JQI.

\end{document}